\documentclass[10pt]{revtex4}
\usepackage{graphicx}
\usepackage{amsmath,amssymb}
\usepackage{mathtools}

\begin{document}

\title{Classical Evolution Without Evolution}
\author{Vlatko Vedral}
\affiliation{Clarendon Laboratory, University of Oxford, Parks Road, Oxford OX1 3PU, United Kingdom\\Centre for Quantum Technologies, National University of Singapore, 3 Science Drive 2, Singapore 117543\\
Department of Physics, National University of Singapore, 2 Science Drive 3, Singapore 117542}

\begin{abstract}
The well known argument of Page and Wootters demonstrates how to ``derive" the usual quantum dynamics of a subsystem in a global state which is an eigenstate of the total Hamiltonian. I show how the same argument can be made in classical physics, by using a formalism that closely resembles the quantum one. This is not surprising since the Hamilton-Jacobi formulation of classical dynamics is precisely motivated by the logic of timeless dynamics. Ultimately, the key to obtaining dynamics without dynamics is the principle of energy conservation which leads to correlations between times pertaining to different subsystems. The same can, of course, be said about all other conserved quantities and we show how to address this problem in its full generality so as to realise rotation without rotation, translation without translation and so on. The classical and quantum interpretations do, however, have one major difference and this is the Church of the Higher Hilbert Space interpretation of mixtures, which only exists in quantum physics. We discuss a few consequences of this point. 
\end{abstract}

\maketitle

We could start this exposition in a number of ways, but the easiest is perhaps to explain how in classical mechanics time could be also thought of as a dynamical variable just like position and momentum. The idea is to start with a time-independent Hamiltonian $H(q,p)$ and construct a new Hamiltonian ${\cal H} = H + p_0$ where $p_0$ is conjugate to the time variable $t$. The equations of motion for this new Hamiltonian (with respect to a new time $\tau$) are: 
\begin{equation}
\frac{\partial {\cal H}}{\partial p_0} = \frac{dt}{d\tau} = 1 \;\;\;  \frac{\partial {\cal H}}{\partial t} = - \frac{dp_0}{d\tau}=0 \; ,
\end{equation}  
from which it follows that $\tau = t+c$ and that $p_0$ is conserved in time. We can set $c=0$ without any loss of generality and can also assume that the value of $p_0$ is such that $H+p_0 = 0$. In other words, the total energy equals zero. We can already see how, at the level of ${\cal H}$, there is no evolution, however, the dynamics generated by the original Hamiltonian remains the same. Namely,
\begin{equation}
\frac{\partial {\cal H}}{\partial p} = \frac{dq}{dt} = \frac{\partial H}{\partial p} \;\;\;  \frac{\partial {\cal H}}{\partial q} = -  \frac{dp}{dt}= \frac{\partial H}{\partial q}\; .
\end{equation}   
This is the formalism of super Hamiltonians (see, for instance, Kramers \cite{Kramers}). It is well known that, when we quantize this system, then $H$ becomes the Hamiltonian operator and $p_0\rightarrow -i\hbar \partial /\partial t$. The resulting equation $(\hat H - i\hbar \partial /\partial t)\psi = 0$ is the Schr\"odinger equation. 

This already tells us that even the time-dependent Schr\"odinger equation can be though of in a timeless way (even Schr\"odinger himself apparently wrote down the time-independent form of his equation first). However, to nudge the classical discussion even closer to the Page-Wootters logic, let us now treat the variables $(p_0,t)$ as actually pertaining to another subsystem, distinct from the one with the conjugate variables $(q,p)$. 

To make the notaion more transparent, the first subsystem will be the one with the variables $(q_1,p_1)$, while the second subsystem will have variables $(q_2,p_2)$. Suppose that the two subsytems are non-interacting which means that the total Hamiltonian is given by $H_1(q_1,p_1)+H_2(q_2,p_2)$. If the total system is closed, the total energy is conserved. We can again assume without loss of generality that the total energy is zero $E=0$, which means that if $H_1(q_1,p_1)=E_1$ and $H_2(q_2,p_2)=E_2$, then $E_1=-E_2$. 

In the Hamilton-Jacobi formalism we thus write:
\begin{equation} 
H(q_1,q_2,\frac{\partial S}{\partial q_1}, \frac{\partial S}{\partial q_2})=E_1+E_2=0\;\; ,
\end{equation}
where $S(q_1,q_2,E) = S_1 (q_1,E_1) + S_2 (q_2,E_2)$ is the total action variable which is sum of the individual actions as the subsystems are non-interacting. Each subsystem could be assigned its own time variable
\begin{equation}
\frac{\partial S_1}{\partial E_1} = t_1 \;\;\;  \frac{\partial S_2}{\partial E_2} = t_2\; .
\end{equation}
Because the total energy is conserved, this immediately implies that $t_1=-t_2$ (we are again omitting any additive constants without loss of generality). Just like in the quantum (Page-Wootters) formalism, the time variables of the two subsystems are correlated, not because the subsystems are coupled, but because the overall energy is conserved. Therefore, we can say that one subsystem ``evolves" dynamically with respect to the other one even though the total system is stationary (i.e. we have that $\partial S/\partial E=0$). 

In the Page-Wootters (PW) work \cite{Wootters}, this relative evolution is immediately transparent through the use of Lie algebras and groups. We first briefly summarize the quantum derivation in order to follow it up with the exact classical analogue.  

In the PW formalism the two subsystems are usually called the system and the clock. The total system is referred to as the universe. The state of the universe is of the form $|\Psi_{cs}\rangle = \sum_t |\phi (t)\rangle|t\rangle$ and it is an eigenstate of the total Hamiltonian (with zero eigenvalue assumed for simplicity), which itself is the sum of the system and the clock Hamiltonians, $H_{sc}=H_s+H_c$. The index $t$ can be thought of as just a dummy index, however, it is clear that it will become the time label of the clock, with respect to which the system states $|\phi (t)\rangle$ will evolve.  We furthermore assume that
\begin{equation}
e^{-iH_s t} |\phi (0)\rangle =|\phi (t)\rangle  \;\;\; e^{-iH_c t} |0\rangle =|t\rangle  \; .
\end{equation}
It is already clear from here that since $e^{-i(H_s + H_c)t}|\Psi_{cs}\rangle=|\Psi_{cs}\rangle$, the dynamics of the system unfolds in sync with the dynamics of the clock, and without any overall dynamics. Namely, if we apply a small time shift to the clock so that the state changes $|t\rangle \rightarrow |t+\delta t\rangle$, then the system will undergo the change $|\phi_t\rangle \rightarrow |\phi (t+\delta t)\rangle$.

Another way of saying the same and obtaining the dynamics of the system is to compute \cite{Marletto}
\begin{equation}
\frac{d |\phi_t\rangle\langle \phi_t|}{dt} = tr_c \Bigg\{ |\Psi_{cs}\rangle \langle \Psi_{cs}| \frac{d |t\rangle\langle t|}{dt}\Bigg\} \; ,
\end{equation}
which, after a sequence of steps, leads us to
\begin{equation}
\frac{d |\phi_t\rangle\langle \phi_t|}{dt} = [ H_s, |\phi_t\rangle\langle \phi_t|] \; ,
\end{equation}
i.e. the Schr\"odinger equation for the system (the bracket $[,]$ is the commutator). 

Can this derivation also be paralelled in classical mechanics? Yes. We can assume that the classical Hamiltonian is, as before, the sum of the system and clock Hamiltonians and that the classical state of the system is a phase space distribution in which the states of the system are correlated to the states of the clock. The dynamics of classical states can be represented as (see e.g. \cite{Sudarshan}):
\begin{equation}
e^{\tilde H (q,p) t} \rho (q,p,0) =\rho (q,p,t) \; ,
\end{equation}
where $H$ is the classical Hamiltonian and  $\rho (q,p,t)$ is the classical phase space distribution at time $t$ (note the absence of the imaginary $i$). $\rho (q,p,t)$ plays the classical role of the state of the system. 

The notation with tilde is a shorthand notation for the following operation
\begin{equation}
\tilde H (\omega)  =\epsilon_{\mu\nu} \frac{\partial H}{\partial x^\mu} \frac{\partial}{\partial \omega^\nu}\; ,
\end{equation}
where $\epsilon_{\mu\nu}$ is the $2k$-dimensional antisymmetric matrix, $\omega$s are the so called unified coordinates that represent all the conjugate variables (the first $k$ coordinates are $q$s and the next $k$ coordinates are conjugate momenta $p$) and $x^{\mu}$ labels the $\mu$-th position coordinate. The Hamilton equations of motion are in this notation given by
\begin{eqnarray}
\frac{d\omega^\nu}{dt} & = &\frac{\partial H}{\partial \omega^{\nu+k}} \;\;\; \nu=1,...k\nonumber\\
\frac{d\omega^\nu}{dt} & = & - \frac{\partial H}{\partial \omega^{\nu-k}} \;\;\; \nu=k+1,...2k \;\;\; .
\end{eqnarray}
The tilde definition is designed so that the Poisson bracket can be written as $\{H (\omega), f(\omega)\} = \tilde H (\omega) f(\omega)$. The iterations of this formula gives us the exponential form of the classical propagator $e^{\tilde H t}$ \cite{Sudarshan}. 

If we now write the Hamiltonian as the sum of two Hamiltonians, so that $\tilde H_{sc} = \tilde H_{s} + \tilde H_{c}$, and assume that the total energy is zero, we will end up reproducing the classical derivation of the evolution of the system relative to the clock. Instead of the Schr\"odinger equation, we will, of course, obtain the Liouville equation in which the commutator will be replaced by the Poisson bracket. 

The group-theoretic formulation of both classical and quantum dynamics therefore helps us expose the analogies as far as the timeless dynamics is concerned. We would now like to discuss one clear difference. Unsurprisingly, it is ultimately related to quantum entanglement. The PW state of the universe is a pure state which corresponds to the maximum knowledge. The reduced states of the system and the clock are mixed, but they are correlated with each other. Classically, as is well known, this is an impossible state of affairs. Classical phase space densities can be correlated, but this automatically implies that their overall state cannot be pure (classically, this means a state with perfectly known positions and momenta of all particles). For instance, we can write to total classical density as
\begin{equation}
\rho_{sc} (\omega_{cs}) = \sum_t p(\omega_{cs}, t) \rho_s (\omega_s, t)\rho_c (\omega_c,t)
\end{equation}
where $p(\omega_{cs}, t)$ is the probability corresponding to the state $t$ (which as in the quantum case could be assumed to be uniform). To each classical state of the clock, there corresponds one classical state of the system.  Formally, $\int d\omega_c \rho_c (\omega_c,t)\rho_{sc} (\omega_{cs}) = \rho_s (\omega_s, t)$, which requires us to assume that the classical states of the clock are ``orthogonal" to each other (meaning that their overlap in phase space is zero as would be the case with two delta functions centred on two different points in phase space). 

In classical physics, the total state (of the universe) cannot be less mixed than the state of any of the subsystems. This presents a problem since, in a classical universe, its analogue PW state can only be explained statistically. All classical mixtures are due to either lack of knowledge or can be seen as describing an ensemble of universes. If the universe is mixed because of the former, would this lack of knowledge be fundamental and why? On the other hand, a statistical ensemble of universes is not a good hypothesis unless we have a strong rationale for introducing it. In quantum mechanics, on the other hand, these problems do no arise. While a mixed state between the system and the clock would also suffice for the PW construction, the point is that quantum physics always allows us to view any mixed state as a pure state in an enlarged Hilbert Space. Quantum universe therefore seems more satisfactory even just from just this formal perspective. An argument could be made, though, that a pure state is somehow special, and that this special state in turn requires an explanation, whereas a maximally mixed state is more natural and follows from some kind of a maximum entropy principle. Either way, it is possibly an additional strength of this approach to time that the exact overall state of the universe may not even be relevant (in the sense that the amplitudes between different states do not have any observable meaning).  

Setting aside the issue of the state of the universe, we would finally like to demonstrate that both classical and quantum discussions can be set in a way that all other conservation laws (other than energy) are included. This is straightforward once it is realised that the Hamiltonian is just one of the relevant generators of transformations. Other generators would be the generators of translations, rotations, boosts and so on whatever is believed to be needed to characterise all the symmetries of the universe. 

Let us enumerate all the relevant generators of displacements as $G^i$, such that say $G^0=H$ and so on. The closed universe consisting of two subsystems would satisfy the following constraint for all $i$, $G^i_s+G^i_c=0$. In other words, not only would the total energy be zero, but so would the total momentum, total angular momentum and so on \cite{MV}. In the analogous way to Page-Wootters, the momentum of one subsystem ``changing" by some value $p$, would ``result" in the other subsystem changing by $-p$ and the same for all other relevant quantities. This can ultimately be done at the level of fields (see e.g. \cite{Weinberg} for a symmetry-based approach to quantum field theory). 

In cosmology, we can think of the two subsystems as being the matter fields (the system) and the gravitational field (the clock) \cite{Banks,Brout,Briggs}. But any quantum evolution that rests on the entanglement between different subsystems can be though of in this way. In quantum physics this takes us back to Mott's analysis of alpha-particle tracks in a cloud chamber \cite{Mott} and ultimately to Everett's relative state interpretation \cite{Everett} of quantum physics. As the present paper shows, all of them could be phrased perfectly well within classical mechanics, with the main difference being that classical correlations invariably imply some kind of a lack of knowledge at the level of the universe. Our work does not take into account the fact that, in general, the subsystems would also interact with one another, and that the total Hamiltonian considered here is therefore only approximate). This raises a number of questions worth investigating in the future.

\textit{Acknowledgments}: The author thanks Chiara Marletto for useful comments. He also acknowledges funding from the National Research Foundation (Singapore), the Ministry of Education (Singapore) and Wolfson College, University of Oxford.


\begin{thebibliography}{99}
%
\bibitem{Wootters} D. Page and W. Wootters, Phys. Rev. D {\bf 27}, 2885 (1983).
%
\bibitem{Kramers} H. A. Kramers, Quantum Mechanics, (Dover Publications 2018). 
%
\bibitem{Marletto} C. Marletto and V. Vedral, Phys. Rev. D {\bf 95}, 043510 (2017).
%
\bibitem{Sudarshan} E.C.G. Sudarshan and N. Mukunda, Classical Dynamics: A Modern Perspective, (Hindustan Book Agency, 2015).
%
\bibitem{Banks} T. Banks, Nucl. Phys. B 249, 332 (1985).
%
\bibitem{Brout} R. Brout, Found. Phys. 17, 603 (1987); R. Brout, G. Horwitz,
and D. Weil, Phys. Lett. B 192, 318 (1987); R. Brout, Z. Phys.
B {\bf 68}, 339 (1987).
%
\bibitem{Briggs} J. S. Briggs and J. M. Rost, Found. Phys. {\bf 31}, 693 (2001). 
%
\bibitem{Mott} N. Mott, Proc. Roy. Soc.  A {\bf 126}, 79 (1929).
%
\bibitem{Everett} H. Everett, {\em On the Foundations of Quantum Mechanics}, (Ph.D. thesis, Princeton University, Department of Physics, 1957).
%
\bibitem{MV} C. Marletto and V. Vedral, The Quantum Totalitarian Property and Exact Symmetries, arxiv.org/abs/2005.00138. 
%
\bibitem{Weinberg} S. Weinberg, The Quantum Theory of Fields (Cambridge University Press 1995). 



\end{thebibliography}
\end{document}